\documentclass[12pt]{article}
\usepackage[margin=1in]{geometry}                % See geometry.pdf to learn the layout options. There are lots.
\usepackage{setspace}
\usepackage{graphicx}
\usepackage{amsmath}
\usepackage{amssymb}
\usepackage{bm}
\usepackage{url}
\usepackage{epstopdf}
\usepackage{times}
\title{Market-based Short-Term Allocations in Small Cell Wireless Networks}
\author{Sayandev Mukherjee and Bernardo A.~Huberman\\ CableLabs\\ \texttt{\{s.mukherjee, b.huberman\}@cablelabs.com}}
\begin{document}
\maketitle

\begin{abstract}
Mobile users (or UEs, to use 3GPP terminology) served by small cells in dense urban settings may abruptly experience a significant deterioration in their channel to their serving base stations (BSs) in several scenarios, such as after turning a corner around a tall building, or a sudden knot of traffic blocking the direct path between the UE and its serving BS. In this work, we propose a scheme to temporarily increase the data rate to/from this UE with additional bandwidth from the nearest Coordinated Multi-Point (CoMP) cluster of BSs, while the slower process of handover of the UE to a new serving BS is ongoing.  We emphasize that this additional bandwidth is \emph{additional} to the data rates the UE is getting over its primary connection to the current serving BS and, after the handover, to the new serving BS.  The key novelty of the present work is the proposal of a decentralized market-based resource allocation method to perform resource allocation to support Coordinated Beamforming (CB) CoMP.  It is scalable to large numbers of UEs and BSs, and it is fast because resource allocations are made bilaterally, between BSs and UEs.  Once the resource allocation to the UE has been made, the coordinated of transmissions occurs as per the usual CB methods.  Thus the proposed method has the benefit of giving the UE access to its desired amount of resources fast, without waiting for handover to complete, or reporting channel state information before it knows the resources it will be allocated for receiving transmissions from the serving BS.
\end{abstract}

\section{Introduction}
Mobile users (or UEs, to use 3GPP terminology) in dense urban settings may abruptly experience a significant deterioration in their channel to their serving base stations (BSs) in several scenarios, such as after turning a corner around a tall building, or a sudden knot of traffic blocking the direct path between the UE and its serving BS.  Although networks are usually planned such that total radio link failure is unlikely before such a UE is either handed over to a new serving BS or a strong connection to the current serving BS is re-established, the UE does experience a sudden and severe drop in data rate for some time.  While this issue has always existed in cellular networks, it takes on new urgency in the age of 5G millimeter wave (mmWave) cells because these cells are small relative to the typical LTE macrocell.  Small cells are traversed more quickly than larger cells, but handover to a single small cell serving BS each time means that there is a high rate of handovers with corresponding control signaling overheads in the network.  Thus, in the context of small cells, it makes sense to have the UE be served not by a single small cell but by a cluster of small cells that transmit to the UE simultaneously, a method called \emph{Coordinated Multi-Point} (CoMP)~\cite{lschmns}.

In this work, we propose to temporarily augment the data rate to and from this UE with a short-term dose of additional bandwidth from the nearest CoMP cluster.  The UE uses this additional bandwidth even as the slower process of handover to a new serving BS is ongoing.  Note that this additional bandwidth is indeed \emph{additional} to the data rates the UE is getting over its primary connection to the current serving BS and, after the handover, to the new serving BS.  When the handover process is complete, the additional bandwidth is expected to be no longer necessary and will be relinquished by the UE.

The key novelty of the present work is the proposal of a decentralized market-based resource allocation method to perform resource allocation to support CoMP.  It is scalable to large numbers of UEs and BSs, and it is fast because resource allocations are made bilaterally, between BSs and UEs.  Once the resource allocation to the UE has been made, the coordinated of transmissions occurs as per the usual CoMP methods.  Thus the proposed method has the benefit of giving the UE access to its desired amount of resources fast, without first waiting for handover to complete, or having to report channel state information in order to know the resources that it will be allocated for receiving transmissions from the serving BS. 

\section{Resource allocation in a CoMP cluster}
A CoMP cluster of BSs can serve UEs in different ways.  A baseline version of CoMP is \emph{coordinated beamforming} (CB)~\cite[Sec.~5.3]{mf} where a BS uses only its own antennas to serve the UEs in its cell, albeit with beamforming across these antennas coupled with coordination across BSs so as to mitigate inter-cell interference.  A more sophisticated CoMP system can perform \emph{joint transmission} (JT)~\cite[Sec.~6.3]{mf} across all BSs in the CoMP cluster, treating all resources (such as bandwidth and antennas) in the CoMP cluster as available to serve all UEs served collectively by the BSs in the cluster. Recently, a third CoMP scheme called \emph{dynamic point selection} (DPS) was introduced as an alternative to handover to support rapid re-routing of data streams as a means to mitigate rapid signal degradation.  In DPS, all coordinating BSs have access to the data streams of all their served UEs, but the specific BS that transmits to that UE can change on a frame-by-frame basis.  This is similar to JT in requiring extensive signaling, communication, and synchronization between the BSs of the CoMP cluster, with the difference from JT being that transmission is by just one BS.

In 3GPP, the process of handover has high latency and imposes a high overhead on the (logical) control signaling channels.  Hence, it is not advisable to do frequent handovers.  Thus, we conceive of a dual-connectivity approach to retaining session quality: after the UE's channel to the present serving BS deteriorates enough to require a handover, we allow the slow and high-overhead handover mechanism to proceed as usual.  However, in order to retain session quality, we will also enable the UE to quickly acquire and aggregate bandwidth from the BSs of the local CoMP cluster in whose service area the UE is present.  The question therefore arises as to how we decide on the relative fraction of resources deployed at each BS in the cluster to support this UE.  

For concreteness, let us consider the case of allocation of just one resource, namely ``bandwidth" (strictly speaking, for an LTE or 5G system, this quantity should be measured in terms of Physical Resource Blocks, or PRBs).

The traditional approach, applied to many kinds of similar problems in various fields of engineering, is to frame the resource allocation problem for a given UE as an optimization problem and solve it at a central controller that handles the coordinated transmissions of the CoMP cluster and therefore has the relevant information on resources in use at each BS in the cluster.  A recent treatment of this approach is given in the ``water-filling" formulation of the resource allocation problem described in~\cite{xjwmp}, especially Example~1 in Section II.C with the quantity $p_k$ being the bandwidth allocation and $a_k$ the spectral efficiency (in bits/s/Hz).  However, if we want to solve this problem simultaneously for multiple UEs, as is likely in the dense urban settings where small cell deployments will exist, the centralized optimization approach does not scale well in terms of computation, storage, or latency.

We note again that this resource allocation from the CoMP cluster is \emph{additional} to the resources the UE is already getting through its connection to the serving BS, which may itself change as a result of the normal handover process.  However, the CoMP resource allocation is designed to complete much faster than the handover from the previous serving BS to the next serving BS (which is probably, though not necessarily, one of the BSs in the CoMP cluster).

\section{Factors hindering conventional CoMP deployment}
JT should be capable of delivering greater gains (as measured in total cell throughput and especially the throughputs to UEs that have poor channels to the strongest BS, called ``cell edge UEs"), but field trials have been disappointing~\cite{idmgfbmtj}. 

JT requires sharing of served UEs' data streams and Channel State Information (CSI) across all BSs in a CoMP cluster that cooperate in JT (called the ``cooperation area"), which imposes strict requirements on timing synchronization and severe loads on the signaling and communication between the BSs of the CoMP cluster.  As summarized recently in~\cite{afgp}, ``these requirements are actually constituting the major downfall of JT CoMP in practical cellular networks, rendering hard to achieve its theoretical gains in practice.  On top of that, ... imperfect and/or outdated CSI and uncoordinated interference have a very large impact on the performance of conventional JT CoMP schemes.  Practical Radio-frequency (RF) components, such as oscillators with phase noise, were also shown to have a similar effect."  

Note that the above issues with deployment of JT because of its high signaling, communications, and synchronization requirements also apply to DPS.  In other words, the problems bedeviling practical deployment of CoMP have remained unchanged for the greater part of a decade, from the time they were enumerated in~\cite[p.~457]{mf}: ``... the importance of having precisely synchronized base station oscillators for downlink joint transmission" and ``... the fact that ... pilot overhead increases linearly in the cooperation size, limits CoMP to scenarios of moderate cooperation size.  Also, ... CoMP gains have to be carefully traded against potentially significant feedback overhead."

Note that CB, which does not promise the high gains of JT but at the same time makes fewer demands on signaling and synchronization than JT or DPS, has been identified as a potential candidate for deployment on both LTE~\cite{afgp} and 5G~\cite{sblj, mr} networks.  However, the tradeoff between CoMP gains and the feedback overhead is a general problem with all CoMP schemes.  The unfortunate truth is that in spite of theoretical promise and a fair amount of hooks in successive 3GPP standards to support it, CoMP has not yet been deployed to a significant extent in any cellular network today.

In the present work, \emph{we will consider only CB}, where the coordinating BSs share the CSI among themselves, but without any need for synchronization. While CB does not require the same heavy signaling loads and stringent synchronization requirements as JT, it is still susceptible to out-of-cluster interference.  However, for the particular application scenario of a CoMP cluster in a high-density urban area with urban canyons being considered here, it is expected that the out-of-cluster interference will be mitigated merely by the presence of the tall buildings and other obstacles to radio wave propagation.  

\section{A new market-based approach to resource allocation}
In the present work, we take a market-based approach to the resource allocation problem, which has the advantage of being scalable to large numbers of UEs and BSs in a CoMP cluster.  As has been pointed out by several researchers (see, for example,~\cite{hjm}), a market-based approach is by definition both decentralized (matching buyers and sellers) and efficient (both buyers and sellers maximize some version of utility and/or profits).  Thus, a market-based approach applied to CoMP resource allocation should be expected to ease some of the signaling load and simplify the synchronization requirements.  

We will discuss two market-based resource allocation schemes to support UEs from a CoMP cluster.  The important common feature of both markets is they are \emph{games} with \emph{strategic actors} (the buyers), i.e., the actions of one buyer influence the actions of other buyers and determine the prices charged by the sellers for the resources sold on the market.  There do exist other market-based frameworks where the buyers are mere price-takers, i.e., they cannot influence the prices charged by the sellers (see for example the PSCP scheme in~\cite{mnra}), but we shall not consider such schemes in the present work.

An early market-based approach to bandwidth assignment (by an MNO) to multiple UEs all using a single application (voice calling) with a small number of distinct quality of service satisfaction levels (QSLs) was proposed in~\cite{icpe} and named ``Bandwidth Market Price" (BMP).  In BMP, each UE has a ``QoS profile" with a possibly different budget for each bandwidth allocation.  Independently and later,~\cite{ba} proposed a scheme named ``BidPacket" with continuously-valued pricing (and corresponding QSLs and budgets) that adapt to the allocated bandwidth, and applicable to many classes of data applications.  BMP may therefore be seen as a special case of BidPacket adapted for voice calling.  We will defer the details of BidPacket to Section~\ref{sec:bidpacket}, as our proposed scheme is a modified version of it.

Several market-based resource allocations have been studied in the context of computing resource allocation to processes and users in a cluster of servers.  The so-called Proportional Sharing scheme (also called Trading Post or Shapley-Shubik Game) was proposed in~\cite{flz}.  Applied to the CoMP scenario, it means that the prospective buyers (UEs) submit bids for the resources, and each UE gets allocated a fraction of the total available resources which is proportional to its bid. A Nash equilibrium was proved to exist in~\cite{flz}, which was then shown in~\cite{bgm} to approximately maximize the Nash social welfare (i.e., the sum of log-throughputs of the UEs).  One advantage of Proportional Sharing is that it can be readily extended to resources of more than one class, e.g., bandwidth and serving BS/antennas in a CoMP cluster.  Unfortunately, the allocation in Proportional Sharing always fully exhausts each UE's budget for the resource, which results in overpayment by UEs (or equivalently, inflated bids for resources).  

A modified Proportional Share with a penalty term was proposed in~\cite{m} to reduce bid inflation by making each bidder pay a cost (to participate in the market) that is proportional to its bid.  It was shown in~\cite{m} that such a scheme has a Nash equilibrium that also maximizes the Nash social welfare.  The scheme in~\cite{m} was simplified and applied to resource allocation in a wireless network in~\cite{ttnphh}\footnote{It appears, however, that the simplified version of~\cite{m} that is proposed in~\cite{ttnphh} has a significant shortcoming, rendering it largely ineffective -- see Appendix~\ref{app:B}.}.  Unfortunately, the iterative allocation algorithm in~\cite{m} requires solving a system of nonlinear equations at each step of the iteration, which is computationally expensive (see Appendix~\ref{app:A}).  Moreover, this system of equations involves the bids from all the UEs.  Therefore, this scheme is better suited for a centralized allocation scheme, say JT, where the solution of the system of equations is done in the CoMP cluster.  We do not discuss Proportional Sharing or its variants in the present work, opting instead to focus exclusively on CB.

\section{Description of the problem}
We now describe the details of the CoMP cluster resource allocation problem, followed by our proposed market-based framework to solve the resource allocation problem.
\begin{enumerate}
\item Each UE gets, with its subscription to the MNO operating the CoMP cluster, a budget to acquire additional bandwidth when needed in the scenario described above.  This budget can be periodically refreshed (say at the start of each MNO billing cycle), or topped up as needed, and leftover budget from the previous billing cycle could be carried over to the next, or converted into a credit toward the MNO's subscription, or into travel miles, vouchers, etc.
\item Suppose UE $i$ is about to turn a corner or do something else that requires a rapid allocation of additional bandwidth in order to maintain session quality.  Say UE $i$ has a budget of $w_i$, which we call its \emph{wealth}.
\item If the quality of UE $i$'s connection to the serving BS or CoMP cluster begins to degrade rapidly during a specified interval of duration $t$, UE $i$ becomes a \emph{buyer} and applies its wealth $w_i$ to purchase bandwidth from the new, local, CoMP cluster.  
\item We assume that a UE can only purchase additional bandwidth from a single CoMP cluster at any given time.  Thus, if UE $i$ was already using additional bandwidth purchased from some CoMP cluster before while served by its serving BS, and now the link to that serving BS and old CoMP cluster has deteriorated enough that the UE needs a rapid allocation of additional bandwidth from a new CoMP cluster to maintain its session, then the bandwidth purchased from the old CoMP cluster is freed up in anticipation of a bandwidth purchase from the new CoMP cluster.
\item Further, each BS in the new CoMP cluster may be viewed as a \emph{seller} of bandwidth on the market defined by the BSs in the CoMP cluster and the UEs that are in the area served by that cluster.  Note that a single UE may purchase bandwidth from more than one BS in the CoMP cluster, and the aggregated bandwidth will be exploited through coordinated transmissions from these sellers.
\item The bandwidth allocation in the above steps to any UE $i$ is only valid for a pre-defined, fixed, short interval (which could, for example, be selected so as to cover the mean time taken to complete the handover of this UE to the next serving BS).  Thus, at the end of this fixed interval, this additional bandwidth that the UE has purchased will be relinquished unless the UE re-enters the market and purchases bandwidth again.
\end{enumerate} 

The following analysis is for a single interval after the rapid deterioration of the channel of an arbitrary UE $i$ to its present serving BS or CoMP cluster has triggered a resource purchase.  A typical scenario for this analysis is when we: (i) predict that UE $i$ will soon need additional bandwidth from a new CoMP cluster and start the timer $T$, (ii) then observe, within the sub-interval of duration $t$ at the beginning of interval $T$, that UE $i$'s channel to its serving BS or old CoMP cluster has worsened by more than some threshold amount, say $\theta$, where $\theta > 0$ is in decibels (dB).  Note also that the analysis applies to bandwidth purchases for transmissions in \emph{a single direction} (i.e., either the \emph{uplink} from UEs to BSs, or the \emph{downlink}, from BSs to UEs).

\section{BidPacket resource allocation}
\label{sec:bidpacket}
BidPacket~\cite{ba} is a market-based bandwidth allocation scheme originally designed for a collection of user devices seeking to transmit on the uplink (i.e., to an access point).  In the original proposal in~\cite{ba}, the sellers and buyers are both WiFi users -- a user with nothing to transmit sells its bandwidth to a user that has a file to transmit and wants more bandwidth than the default allocation (which is the same for all users).  In the present work, the buyers are the UEs, and the sellers are the BSs of the CoMP.

\subsection{Utility model for UEs}
Let $p$ be the price per unit of bandwidth on the bandwidth market comprising the new CoMP cluster BSs as the sellers, and the UEs in the service area of this CoMP cluster as the buyers.  We employ the buyer utility function proposed in~\cite{ba}: if UE $i$ purchases bandwidth $B_i$, its utility is
\begin{equation}
	U_i = b_i B_i - \frac{1}{2 w_i} p B_i^2,
	\label{eq:utility_i}
\end{equation}
where $w_i$ is the \emph{wealth} of UE $i$, and $b_i \leq 1$ is a measure of the \emph{need} of UE $i$ for bandwidth. 

For example, suppose UE $i$'s channel to its serving BS or old CoMP cluster at the end of the sub-interval of length $t$ is $\tau_i > \theta$ (both $\tau$ and $\theta$ are in dB) \emph{worse} than at the beginning of this sub-interval.  Say $\tau_{\max}$ (in dB) is the maximum deterioration in the channel to the serving BS or old CoMP cluster that can be tolerated before the session is interrupted.  Then we could define $b_i$ to be the ratio $\tau_i/\tau_{\max}$ in the linear scale, i.e., $b_i = 10^{(\tau_i-\tau_{\max})/10}$.  In other words, the greater the deterioration of the UE's channel to its serving BS or old CoMP cluster, the greater its need for additional bandwidth in order to maintain the session, and the greater the utility it derives from a bandwidth purchase from sellers in the new CoMP cluster.

As shown in~\cite{ba}, UE $i$ maximizes its utility by spending the fraction $b_i$ of its wealth $w_i$ to purchase bandwidth, i.e., by purchasing an amount of bandwidth $B_i$ given by
\begin{equation}
	p B_i = w_i b_i.
	\label{eq:bmp}
\end{equation}
Note that the price paid by UE $i$ for acquiring bandwidth $B_i$ is $p B_i$.  Now, paying by bandwidth is not the usual pricing scheme in cellular networks today.  MNOs either charge a monthly subscription or price by data usage.  The latter is more appropriate for our scenario, since these bandwidth purchases are for short durations of time defined by allocation epochs.  In the above, the price $p$ is actually the price per unit of bandwidth per allocation epoch.  

If the allocation epoch is a unit of time, and $S_i$ is the average spectral efficiency (throughput per unit of bandwidth used) to UE $i$ over that unit of time, then the total data to or from UE $i$ over that unit of time is $T_i = B_i S_i$.  Thus, in conventional terms, the charge levied by the MNO for the data transmitted to or from UE $i$ during the allocation period may be seen as $\tilde{p} T_i = \tilde{p} S_i B_i = p B_i$, where the price per unit of data is $\tilde{p} = p/S_i$, the bid price $p$ paid by the UE per unit of bandwidth, divided by the average spectral efficiency $S_i$ over the allocation period.  Note that $\tilde{p}$ is precisely the BMP of~\cite{icpe}.  Moreover,~\eqref{eq:bmp} shows that $p$ is inversely proportional to the demanded bandwidth, exactly as in the QoS profile proposed in~\cite{icpe} without the restriction to finitely many QoS satisfaction levels.  Thus, the BMP scheme of~\cite{icpe} may be seen as a special case of BidPacket.

\subsection{Profit model for BSs in the CoMP cluster}
Each BS in the CoMP cluster, being a seller of bandwidth, seeks to maximize its profit from bandwidth sales.  Although the BSs (or more precisely, the MNOs operating these BSs) have already paid a fixed price (at an FCC spectrum auction) for the bandwidth that they are selling, it is prudent for each BS not to seek to sell all of its available bandwidth all the time, but to conserve the amount of total bandwidth it sells, i.e., minimize the total bandwidth in use at this BS.  This way, the BS could cope with a sudden surge of demand arising from a spike in traffic caused, for example, by an influx of UEs into the service area of this BS and CoMP cluster.

Therefore we shall use the cost function defined in~\cite{ba}: the cost of selling bandwidth $B^{(j)}$ for BS $j$ in the CoMP cluster is 
\begin{equation}
	C(B^{(j)}) = \frac{\left(B^{(j)}\right)^2}{2 a_j},
\end{equation}
where $a_j$ is a measure of the importance of conserving bandwidth at BS $j$.  Note that $C(B^{(j)})/B^{(j)}$ increases with $B^{(j)}$, which means that the cost per unit of bandwidth increases with the bandwidth.

The profit to BS $j$ from selling bandwidth $B^{(j)}$ on the bandwidth market is therefore the difference between its revenue and its cost:
\begin{equation}
	\rho_j = p B^{(j)} - C(B^{(j)}),
\end{equation}
and, as shown in~\cite{ba}, the BS's profit is maximized by selling the amount of bandwidth given by
\begin{equation}
	B^{(j)} = a_j p.
\end{equation}

\subsection{Equilibrium pricing on the bandwidth market}
The utility-maximizing total demanded bandwidth from all UEs in the service area of this CoMP cluster is
\begin{equation}
	B_{\mathrm{demand}} = \sum_{\text{UEs }i} B_i = \frac{1}{p}\sum_{\text{UEs }i} w_i b_i,
\end{equation}
and the profit-maximizing total supplied bandwidth from all BSs in this CoMP cluster is
\begin{equation}
	B_{\mathrm{supply}} = \sum_{\text{BSs }j} B^{(j)} = p\sum_{\text{BSs }j} a_j.
\end{equation}
It follows that at equilibrium, the price per unit of bandwidth on the market is such that the bandwidth supply equals the bandwidth demand~\cite{ba}:
\begin{equation}
	p = \sqrt{\frac{\sum_{\text{UEs }i} w_i b_i}{\sum_{\text{BSs }j} a_j}}.
	\label{eq:p_equi}
\end{equation}

Recall from~\eqref{eq:bmp} that at the price~\eqref{eq:p_equi}, each UE $i$ will purchase an amount $B_i$ of bandwidth such that the total price it pays is $w_i b_i$.  In other words, at equilibrium, $w_ib_i$ is precisely the value of UE $i$'s \emph{bid} for bandwidth.  Thus we may rewrite~\eqref{eq:p_equi} as
\begin{equation}
	p = \sqrt{\frac{\sum_{\text{UEs }i} \text{bid}_i}{\sum_{\text{BSs }j} a_j}}.
	\label{eq:p_equi2}
\end{equation}

Note that BidPacket buyers' budgets may be funded with virtual currency, but they should be linked to some currency or credit (like airline miles or discount coupons) with monetary value in the real world.  Otherwise, with a purely virtual currency with no real-world value, it is optimal for the buyers to spend their entire budgets every time for bandwidth purchase, and to overstate their urgency/need in order to spend their entire budget.
 
In the CoMP scenario, the buyers are the UEs, as stated earlier, and the sellers (under the assumption of coordinated beamforming) are the BSs in the CoMP cluster. The BidPacket scheme is in the form of transactions between individual pairs of buyers and sellers.  It thus has the advantage of maximizing both buyer and seller utility at equilibrium, while being scalable to large numbers of buyers and sellers.    

BidPacket is really only applicable to a single resource (like bandwidth) whereas in a CoMP cluster the available resources are multi-dimensional (like bandwidth and antenna selection, for example). Lastly, there are no theoretical results on whether or not a Nash Equilibrium exists among the UEs such that no UE can change its bids to improve its utility without decreasing the utility of another UE.

\section{Bandwidth allocation algorithm}
\begin{enumerate}
\item At the start of each epoch, (a) the BSs in the CoMP cluster get assigned a random order for serving UE requests for bandwidth purchases; (b) the UEs submit their total bid amounts to a bid table that is accessible to all BSs in the CoMP cluster.
\item Following the order assigned to the BSs, the UEs' bandwidth requests are served by those BSs, starting from the UE with the highest bid, then the UE with the next highest bid, and so on.  If a UE's bandwidth request is too much to be satisfied by a single BS, the next BS satisfies the remaining part of the request.  When a BS satisfies the last remaining part of a UE's request, that UE's bid is removed from the bid table.
\item All BSs that together satisfy a UE's bandwidth request now coordinate their transmissions to that UE in the next epoch, following conventional CoMP protocols.
\end{enumerate}
Note that the algorithm requires only a single pass through all UEs requesting bandwidth and all BSs that provide that bandwidth.

\section{Numerical results}
We simulate a scenario where $10$ UEs are bidding for bandwidth from a CoMP cluster of $4$ BSs.  Any UE can purchase bandwidth from any of the $4$ BSs.  Each BS has $25$ units of bandwidth, so that the total bandwidth available in the CoMP cluster is $100$ units. In this simple scenario we further assume perfect beamforming, so any bandwidth can be used for simultaneous transmissions by multiple BSs.  Each UE starts with an initial budget of $C$ units in some virtual currency (where $C=500, 1000, 5000$), and a default bandwidth allocation of $100/10=10$ units.  

Note that once a bandwidth assignment is made, the actual transmissions are exactly those of conventional CB CoMP, hence we have simulated only the bandwidth assignments themselves.

The simulation setup follows that in~\cite{ba}: at each allocation epoch, a UE wants to receive, with fixed probability, a video file with length modeled by a Gaussian random variable with mean $150$ units and standard deviation $50$ units.  For simplicity, we assume that one unit of file length requires a single transmission over one unit of bandwidth, and we ignore any channel imperfections or possibility of packet error.  A UE can either use the default bandwidth $10$ that it has been originally assigned, or purchase more bandwidth for a certain amount of time as per the utility function described above.  In the simulation of the utility, the need $b_i$ for bandwidth at UE $i$ is drawn from the uniform distribution on the interval $(0,1)$. Similarly, the quantity $a_j$ for each BS $j$ is also drawn from the uniform distribution on $(0,1)$, and all $a_i$ and all $b_j$ are independent.

\begin{figure}[htbp]
\begin{center}
\includegraphics[width=\linewidth]{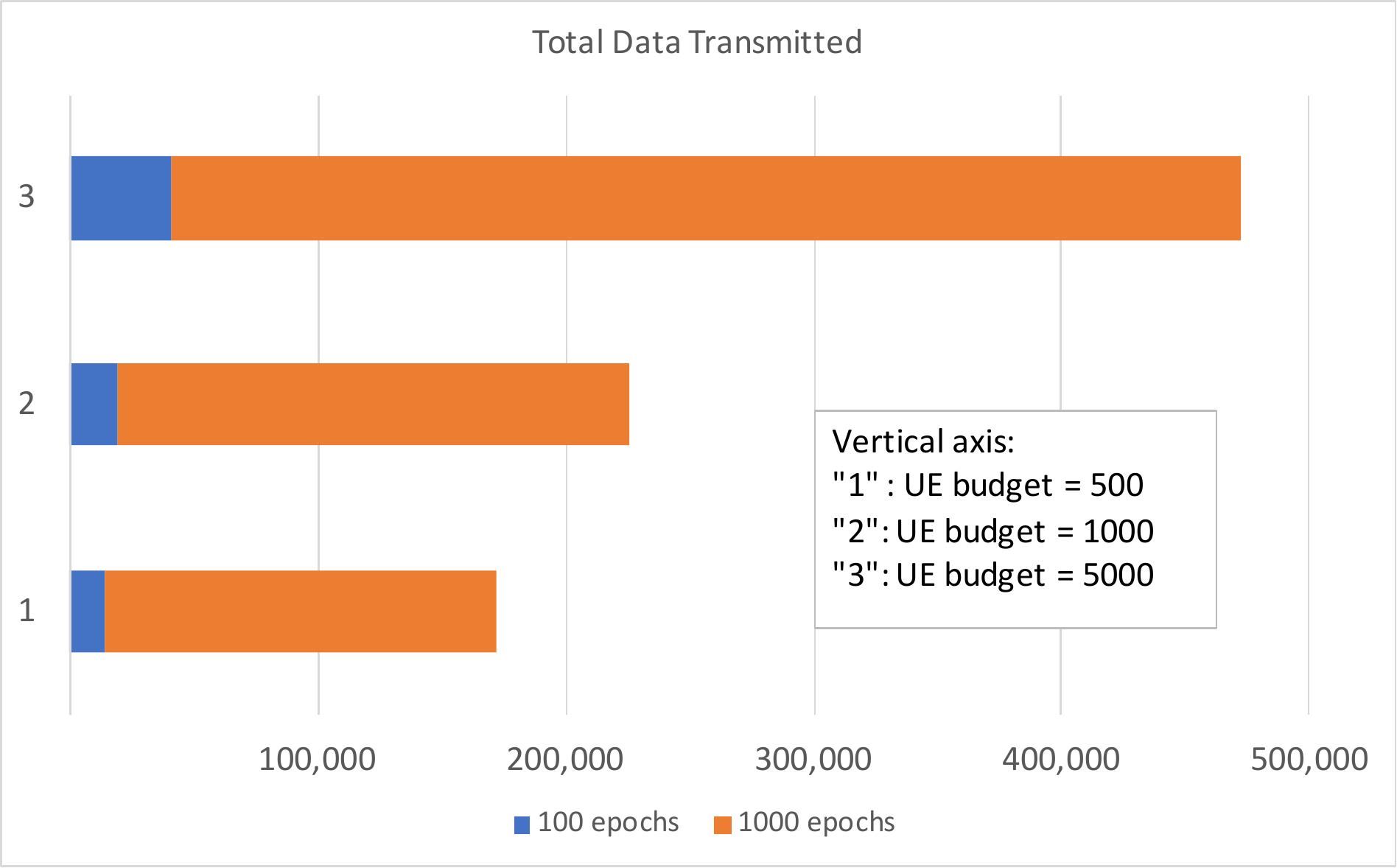}
\caption{Plot of total data transmitted versus total overall bandwidth, for the BidPacket bandwidth purchase strategy and the baseline with fixed allocation of bandwidth to each UE.}
\label{fig:bandwidth}
\end{center}
\end{figure}

Fig.~\ref{fig:bandwidth} is a plot of the total data transmitted (in terms of the above units) over $100$ and $1000$ time periods by the $10$ UEs for the three different values of each UE's initially assigned budget $C$, under the market described above versus the baseline non-market scenario when UEs cannot purchase additional bandwidth and must use their default bandwidth of $10$ units.  Note that no UE's budget is replenished during $100$ or $1000$ epochs over which the throughput is aggregated.  Thus, if a UE exhausts its budget after a certain number of epochs, it will have to fall back on its default bandwidth for subsequent epochs.  

For comparison, note that with the default allocation of bandwidth of $10$ units per UE, the total data over $100$ epochs is $10,000$ whereas over $1000$ epochs the total data is $100,000$. It is clear that the market-based bandwidth purchasing significantly increases the total data transmitted over the baseline which does not permit the UEs to purchase additional bandwidth.

For the small budget of $C=100$, we observe that even for $100$ epochs, the total data is actually less than that with the default allocation of bandwidth, meaning that the UEs exhaust their budgets earlier than $100$ epochs.  However, with a relatively modest increase of budget per UE from $100$ to $500$, we observe that each UE has nonnegative budget at the end of even $1000$ epochs.  Thus in this simple scenario the concerns about inflated bids with virtual currency do not apply.

\appendix

\section{Overview of allocation scheme in~\cite{m}}
\label{app:A}
In~\cite{m}, a variant of Proportional-Share allocation with a penalty term is proposed and analyzed.  Suppose there are a total of $R$ units of bandwidth, and $n$ UEs with bids\footnote{Note that we are now using the notation $b$ for a \emph{bid} rather than for the \emph{need/urgency} as in Section~\ref{sec:bidpacket}.} $\bm{b} = [b_1,\dots,b_n]^{\mathrm{T}}$.  The penalty term is proportional to the total bid amount, i.e., the utility function of UE $i$ is 
\[
	u_i(\bm{b}, q_i) = v_i(r_i(\bm{b})) - q_i b_i, \quad i=1,\dots,n,
\]
where $r_i(\bm{b})$ is the assigned bandwidth to UE $i$ under Proportional-Share, i.e., $r_i(\bm{b}) = R b_i/(b_1+\cdots+b_n)$, $v_i(\cdot)$ is the \emph{valuation function} for UE $i$ (see below for details), and $q_i$ is the \emph{cost} (per unit bid amount) to UE $i$ to participate in the auction, and is set by the seller.

The valuation function $v_i(r)$ for UE $i$ is the logarithm of the data rate to $i$ when it is served by the CoMP cluster with bandwidth $r$:
\begin{equation}
	v_i(r) = \ln(1 + r\,\mathrm{SE}_i), \quad \mathrm{SE}_i = \log_2\left(1 + \frac{P_i H_i}{N_0}\right),
	\label{eq:v}
\end{equation}
where $\mathrm{SE}_i$ is the spectral efficiency on the downlink to UE $i$, and is given by the Shannon Formula~\eqref{eq:v}, where $P_i$ is the downlink transmit power of the CoMP cluster to UE $i$ (from one or several serving BSs in the CoMP cluster) per unit of bandwidth, $H_i$ is the channel gain of UE $i$, and $N_0$ is the noise power spectral density. 

The following results are proved in~\cite{m}:
\begin{enumerate}
\item For $n>1$, any strictly positive resource assignment $\bm{r} = [r_1,\dots,r_n]^{\mathrm{T}}$ can be obtained as the Proportional-Share assignment of the unique Nash equilibrium (NE)\footnote{At NE, for these bid amounts and penalties, no UE can unilaterally change its bid in order to improve its utility.} $[\tilde{b}_1,\dots,\tilde{b}_n]^{\mathrm{T}}$ for some set of penalties $\bm{q} = [q_1,\dots,q_n]^{\mathrm{T}}$, which are themselves unique when normalized by their sum~\cite[Thm.~5]{m}.  
\item The penalties $\bm{q}^*$ at the NE yielding the Proportional-Share assignment $\bm{r}^*$ that optimizes the social welfare~\cite[Thm.~8]{m}
\begin{equation}
	\arg\max_{\bm{r}^*}\left\{\sum_{i=1}^n v_i(r_i) \Bigg| \sum_{i=1}^n r_i \leq R, \quad r_i \geq 0, \ i=1,\dots,n\right\}
	\label{eq:maxsw}
\end{equation}
are given by the following indirect expression~\cite[eqn.~(13)]{m}:
\begin{equation}
	r_i^* = R\left[1 - \frac{(n-1) q_i^*}{\sum_{j=1}^n q_j^*}\right], \quad i=1,\dots,n.
	\label{eq:rqstar}
\end{equation}
\item The above $\bm{q}^*$ can be found as follows: from any initial $\bm{q}(0)$, the price trajectory $\bm{q(t)}$ governed by the differential equation
\begin{equation}
	\frac{\mathrm{d}}{\mathrm{d}t} q_i(t) = \frac{R - r_i(t)}{n-1} - \frac{R q_i(t)}{\sum_{j=1}^n q_j(t)}, \quad i=1,\dots,n,
	\label{eq:qde}
\end{equation}
converges to $\bm{q}^*$ as $t \to \infty$~\cite[Thm.~9]{m}, where $r_i(t) \equiv r_i(\bm{q}(t))$, $i=1,\dots,n$, the NE allocations under penalty $\bm{q}$ at time $t$, are the solutions to the system of $n$ equations~\cite[eqn.~(7), Thm.~2]{m}
\begin{align}
	\frac{[R - r_1(t)]v_1'(r_1(t))}{q_1(t)} = \frac{[R - r_2(t)]v_2'(r_2(t))}{q_2(t)} = \cdots = \frac{[R - r_n(t)]v_n'(r_n(t))}{q_n(t)}, \label{eq:ne1} \\
	r_1(t) + r_2(t) + \cdots + r_n(t) = R. \label{eq:ne2}
\end{align}
\end{enumerate}

In practice, we change~\eqref{eq:qde} to the following discrete version: at the $k$th iteration, update:
\begin{equation}
	q_i^{(k)} = q_i^{(k-1)} + \delta\left(\frac{R - r_i^{(k-1)}}{n-1} - \frac{R q_i^{(k-1)}}{\sum_{j=1}^n q_j^{(k-1)}}\right),
	\label{eq:q}
\end{equation}
where $\delta$ is a small positive step size.  Let $x_i = (R-r_i^{(k)}) v_i'(r_i^{(k)})$, $i=1,\dots,n$.  From~\eqref{eq:ne1}, we have
\[
	\frac{x_1}{q_1^{(k)}} = \cdots = \frac{x_n}{q_n^{(k)}} = \frac{X}{\sum_{j=1}^n q_j^{(k)}} \Rightarrow  x_i = \rho_i X, \quad \rho_i = \frac{q_i^{(k)}}{\sum_{j=1}^n q_j^{(k)}}, \quad i=1,\dots,n, \quad X = \sum_{j=1}^n x_j,
\]
and from~\eqref{eq:v}, we have
\begin{equation}
	r_i^{(k)} = \frac{R - x_i / \mathrm{SE}_i}{1 + x_i} = \frac{R - \rho_i X / \mathrm{SE}_i}{1 + \rho_i X}, \quad i=1,\dots,n,
	\label{eq:rX}
\end{equation}
so substituting in~\eqref{eq:ne2} yields the following polynomial equation for $X$:
\begin{equation}
	\frac{R - \rho_1 X / \mathrm{SE}_1}{1 + \rho_1 X} + \cdots + \frac{R - \rho_n X / \mathrm{SE}_n}{1 + \rho_n X} = R.
	\label{eq:X}
\end{equation}
Note that at each step $k$ of the iteration above, we have to solve the polynomial equation~\eqref{eq:X} that in general is of degree $n$ in $X$, for a real root $X$. In general, the complexity of finding the roots of an $n$th-degree polynomial is $O(n^2)$ Boolean (bitwise) operations~\cite[Thm.~7]{pz}. Not only must this computational burden be borne by the CoMP cluster, because only it knows all the terms of~\eqref{eq:X}, but also the root-finding algorithm is iterative, requiring $d$ iterations to approximate the real roots to an accuracy of about $2^{-d}$ \emph{at each step} $k$ of the allocation algorithm~\eqref{eq:q},~\eqref{eq:X},~\eqref{eq:rX}.

\section{Problems with the iterative algorithm proposed in~\cite{ttnphh}}
\label{app:B}
In~\cite{ttnphh}, an iterative algorithm is proposed for resource allocation that seemingly avoids the need to solve the $n$th-degree polynomial equation~\eqref{eq:X} at each step of the iteration as required by~\cite{m}.  However, as we shall show below, the algorithm in~\cite{ttnphh} creates a circular sequence of updates that leads to very undesirable outcomes.
\subsection{Iterative algorithm for bidding and allocation under Proportional-Share with penalty}
\label{sec:algo1}
The iterative algorithm in~\cite[Algorithm~1]{ttnphh} starts by setting $q_i^{(0)}$ to some small value, $\mu_i^{(0)} = 0$, an initial assignment of resources $r_i^{(0)}$, $i=1,\dots,n$, and initial bids calculated as follows:
\begin{equation}
	b_i^{(0)} = \frac{1}{q_i^{(0)}} r_i^{(0)} v_i'(r_i^{(0)}) [1-\mu_i^{(0)}] = \frac{1}{q_i^{(0)}} r_i^{(0)} v_i'(r_i^{(0)}), \quad i=1,\dots,n.
	\label{eq:binit}
\end{equation}
Subsequently, at iteration $k$, we make the following updates in the order written:
\begin{align}
	\mu_i^{(k)} &= 1 - \frac{b_i^{(k-1)} q_i^{(k-1)}}{r_i^{(k-1)} v_i'(r_i^{(k-1)})}
	\label{eq:mu} \\
	q_i^{(k)} &= q_i^{(k-1)} + \delta\left(\frac{R - r_i^{(k-1)}}{n-1} - \frac{R q_i^{(k-1)}}{\sum_{j=1}^n q_j^{(k-1)}}\right), \tag{\ref{eq:q}}\\
	b_i^{(k)} & = \frac{1}{q_i^{(k)}} r_i^{(k-1)} v_i'(r_i^{(k-1)})[1 - \mu_i^{(k)}],
	\label{eq:b} \\
	r_i^{(k)} &= R \frac{b_i^{(k)}}{\sum_{j=1}^n b_j^{(k)}}.
	\label{eq:rps}
\end{align}
A key observation is that the updates~\eqref{eq:mu} and~\eqref{eq:b} are circular.  Its undesirable consequence is that \emph{the final assignment depends only on the (random) initial assignment}.  We prove this below.

First, we note that from~\eqref{eq:mu}, we have for all $k=1,2,\dots$,
\begin{equation}
	r_i^{(k-1)} v_i'(r_i^{(k-1)}) [1 - \mu_i^{(k)}] = b_i^{(k-1)} q_i^{(k-1)}, \quad i=1,\dots,n.
	\label{eq:constant}
\end{equation}
Applying~\eqref{eq:constant} in~\eqref{eq:b}, we then have
\[
	b_i^{(k)}q_i^{(k)} = r_i^{(k-1)} v_i'(r_i^{(k-1)}) [1 - \mu_i^{(k)}] =  b_i^{(k-1)} q_i^{(k-1)},
\]
which when applied repeatedly yields
\begin{equation}
	b_i^{(k)}q_i^{(k)} = b_i^{(k-1)} q_i^{(k-1)} = \cdots = b_i^{(0)} q_i^{(0)} = r_i^{(0)} v_i'(r_i^{(0)}) = c_i, \text{ say}, \quad i=1,\dots,n, \label{eq:const}
\end{equation}
where in the final step we have used~\eqref{eq:binit}.  

It follows from~\eqref{eq:const} that at each iteration $k$, the resource allocation to UE $i$ is given by
\begin{equation}
	r_i^{(k)} = R \frac{b_i^{(k)}}{\sum_{j=1}^n b_j^{(k)}} = R\frac{c_i/q_i^{(k)}}{\sum_{j=1}^n c_j/q_j^{(k)}} = R\frac{c_i/q_i^{(k)}}{H^{(k)}}, \quad i=1,\dots,n, \qquad H^{(k)} = \sum_{j=1}^n \frac{c_j}{q_j^{(k)}}.
	\label{eq:r}
\end{equation}
Thus the allocation algorithm~\eqref{eq:binit}--~\eqref{eq:rps} can be written in the following mathematically equivalent form: start by initializing $q_i^{(0)}$ to some small value, as before.  At each iteration $k$, update~\eqref{eq:q}:
\begin{equation}
	q_i^{(k)} = q_i^{(k-1)} + \delta \frac{R}{n-1}\left(1 - \frac{c_i/q_i^{(k-1)}}{H^{(k-1)}} - \frac{q_i^{(k-1)}}{Q^{(k-1)}}\right), \quad Q^{(k-1)} = \frac{1}{n-1}\sum_{j=1}^n q_j^{(k-1)}
	\label{eq:newq}
\end{equation}
until convergence to $q_i^*$, $i=1,\dots,n$.  From~\eqref{eq:r}, the final resource assignments are
\[
	r_i^* = R\frac{c_i/q_i^*}{H^*}, \quad i=1,\dots,n, \qquad H^* = \sum_{j=1}^n \frac{c_j}{q_j^*}.
\]
From~\eqref{eq:newq} and~\eqref{eq:constant} it follows that the final assignments $r_i^*$ depend \emph{only on} $c_i = r_i^{(0)} v_i'(r_i^{(0)})$, $i=1,\dots,n$.  This is completely undesirable because it means the algorithm deterministically yields final assignments depending \emph{only} on the random \emph{initializations} to the iterative algorithm.

\end{document}